\documentclass[a4paper,12pt]{article}
\usepackage{graphicx}
\usepackage{infnprep}
\usepackage[top=2.5cm, bottom=3cm, left=3cm, right=3cm]{geometry}

\newcommand{\Header}{
  \resizebox{15cm}{!}{
  \begin{tabular}{rl}
  \includegraphics[width=5cm, trim={50 100 0 0}]{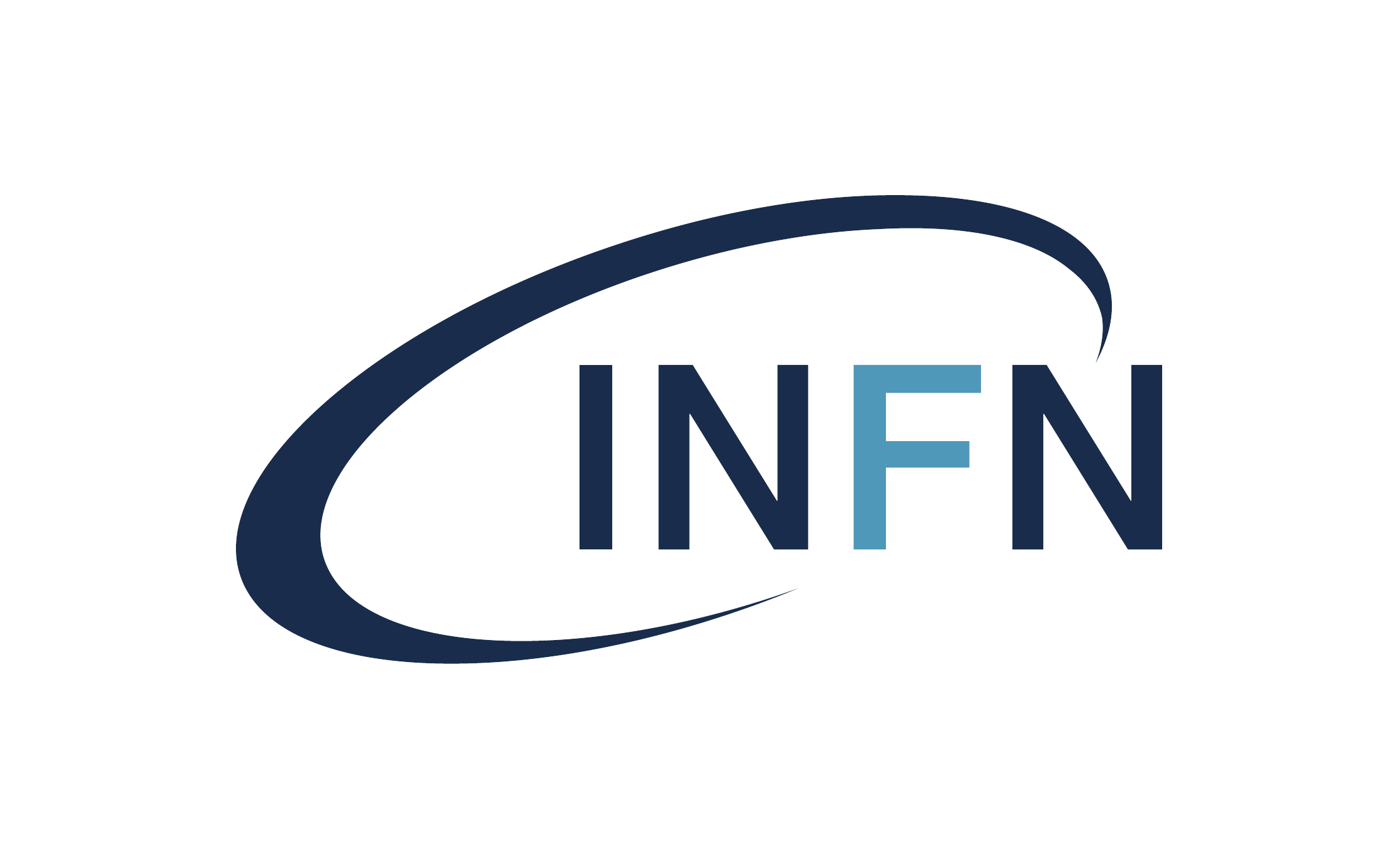} & {\LARGE\sffamily ISTITUTO NAZIONALE DI FISICA NUCLEARE}\\
      \\
  \end{tabular}
  }
    \renewcommand{\arraystretch}{1}
\vskip 0.5cm
\rule{15.0cm}{0.09mm}
  \begin{flushright}
      {\underline{\bf INFN-19/2/LNF}}\\    
      {\small\bf 11 January 2019} \\      
  \end{flushright}
}

\begin{document}
\begin{titlepage}
\title
  {\Header \large \bf CYGNO: a CYGNUs Collaboration 1 m$^3$ Module with Optical Readout for Directional Dark Matter Search}
%

\author{CYGNO Collaboration}

\maketitle

\begin{abstract}
The design of the project named CYGNO is presented. CYGNO is a new proposal supported by INFN, the Italian National Institute for Nuclear Physics, within CYGNUs proto-collaboration (CYGNUS-TPC) that aims to realize a distributed observatory in underground laboratories for directional Dark Matter (DM) search and the identification of the coherent neutrino scattering (CNS) from the Sun. CYGNO is one of the first prototypes in the road map to 100-1000 m$^3$ of CYGNUs and will be located at the National Laboratory of Gran Sasso (LNGS), in Italy, aiming to make significant advances in the technology of single phase gas-only time projection chambers (TPC) for the application to the detection of rare scattering events. In particular it will focus on a read-out technique based on Micro Pattern Gas Detector (MPGD) amplification of the ionization and on the visible light collection with a sub-mm position resolution sCMOS (scientific COMS) camera. This type of readout - in conjunction with a fast light detection - will allow on one hand to reconstruct 3D direction of the tracks, offering accurate sensitivity to the source directionality and, on the other hand, a high particle identification capability very useful to distinguish nuclear recoils. 
\end{abstract}

\begin{flushleft}
\end{flushleft}
\begin{flushright}
\small\it Published by \\
Laboratori Nazionali di Frascati\\
Conference Record of 2018 IEEE NSS/MIC/RTSD\\
presented by G.Mazzitelli
\end{flushright}
\end{titlepage}
\pagestyle{plain}
\setcounter{page}2
\baselineskip=17pt

\section{Introduction}
\begin{figure}[!ht]
\centering
\includegraphics[width=0.9\linewidth, angle=0]{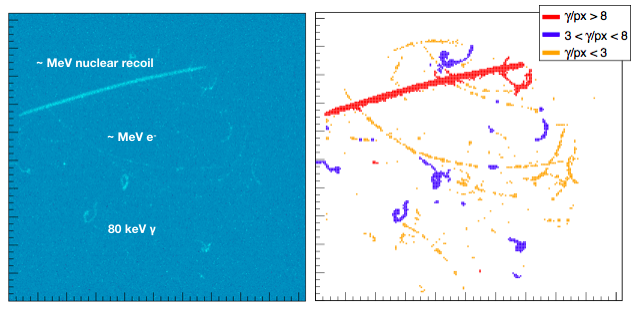}

\caption{Particle identification in ORANGE prototype based on HeCF$_4$ optically readout  - left: example of raw image collected with AmBe source neutron recoil and background (e$^-$, $\gamma$); right: PID clustering track analysis reconstruction}
\label{fig:pid}
\end{figure}

Many ton-scale and multi-ton scale underground experiments based on dual phase TPCs liquid Ar or Xe as active target are exploring WIMP masses ($M_{\chi}$) larger than 10 GeV, while lighter target nuclei can be exploited as alternative targets to extend the searches in the range of lower DM mass, down to few GeV mainly through small solids mass cooled down to very low temperature. CYGNO\footnote{
{CYGNO Collaboration: E. Baracchini$^{2,15}$, R. Bedogni$^{1}$, F. Bellini$^{3,5}$, L. Benussi$^{1}$, S. Bianco$^{1}$, L. Bignell$^{4}$, M. Caponero$^{1, 12}$, G. Cavoto$^{3,5}$, E. Di Marco$^{5}$, C. Eldridge$^{6}$, A. Ezeribe$^{6}$, R. Gargana$^{1}$, T. Gamble$^{6}$, R. Gregorio$^{6}$, G. Lane$^{4}$, D. Loomba$^{7}$, W. Lynch$^{6}$, G. Maccarrone$^{1}$,  M. Marafini$^{5,8}$, G. Mazzitelli$^{1}$,  A. Messina$^{1,5}$, A. Mills$^{7}$, K. Miuchi$^{10}$, F. Petrucci$^{11,13}$, D. Piccolo$^{1}$, D. Pinci$^{5}$, N. Phan$^{7}$, F. Renga$^{5}$, G. Saviano$^{1,14}$, 1, N. Spooner$^{6}$, T. Thorpe$^{9}$, S. Tomassini$^{1}$, S. Vahsen$^{9}$\\ 
\textit{$^{1}$ Istituto Nazionale di Fisica Nucleare, Laboratori Nazionali di Frascati, Frascati (RM), Italy}\\
\textit{$^{2}$ Gran Sasso Science Institute, L'Aquila, Italy}\\
\textit{$^{3}$ Sapienza Università di Roma, Dipartimento di Fisica, Rome, Italy}\\
\textit{$^{4}$  Australian National University, ACT, Canberra, Australia}\\
\textit{$^{5}$  Istituto Nazionale di Fisica Nucleare, Sezione di Roma, Rome, Italy}\\
\textit{$^{6}$  University of Sheffield, Sheffield, United Kingdom}\\
\textit{$^{7}$  University of New Mexico, Albuquerque, New Mexico, United States of America}\\
\textit{$^{8}$  Museo Storico della Fisica e Centro Studi e Ricerche Enrico Fermi, Rome, Italy}\\
\textit{$^{9}$  University of Hawaii, Honolulu, Hawaii, United States of America}\\
\textit{$^{10}$  Kobe University, Kobe, Japan}\\
\textit{$^{11}$  Istituto Nazionale di Fisica Nucleare, Sezione di Roma TRE, Rome, Italy}\\
\textit{$^{12}$  ENEA, Frascati, Frascati (RM), Italy}\\
\textit{$^{13}$  Università di Roma TRE, Dipartimento di Matematica e Fisica, Rome, Italy}\\
\textit{$^{14}$  Sapienza Università di Roma, Dipartimento di Ingegneria Chimica Materiali Ambiente, Rome, Italy}\\
\textit{$^{15}$  Istituto Nazionale di Fisica Nucleare, Laboratori Nazionali del Gran Sasso, L'Aquila, Italy}}
} aims to access a middle energy and mass range by building a gas-only TPC chamber that can be operated at room temperature and atmospheric pressure based on a gas mixture of He and CF$_4$~\cite{NIM:Fragaetal}: the particles recoiling after a scattering in the active gas volume leave a trail of ionization that is drifted to the anode where it is amplified by means of a Micro Pattern Gas Detector (MPGD) like a triple Gas Electron Multiplier (GEM)~\cite{NIM:Margatoetal}. Hence, a visible scintillation light is emitted in the electron amplification process and partially collected by a fast detector (PMT/SiPM) and  focused on a sCMOS camera featuring high granularity and extremely low noise per pixel (Fig.~\ref{fig:pid})~\cite{NIM:Marafinietal}, ~\cite{JINST:Marafinietal}, ~\cite{IEEE:Mazzitelli2017}, ~\cite{Pinci:2017goi}, ~\cite{Marafini:2016kzd}, ~\cite{Antochi:2018otx}, ~\cite{nim:pincielba}.

This gas-only detector has a low density (kg/m$^3$) compared to liquid or solid target (ton/m$^3$), calling for large volumes to be competitive, it can however feature a very low detection threshold (down the level of the gas molecule ionization potential) and new ways to reject backgrounds, based on the ionization density or the topological information present in the recoiling particle image. Moreover, this can be used to infer the DM original direction and to access to solar neutrino floor, feature hardly accessible to small solid detectors, as well as, large liquid detectors.

A ton gas detector can be realized, thanks to optical readout of MPGD, with a relatively low cost technology and easy to be scaled to the required volume. The aim of CYGNO - in the context of the CYGNUS international collaboration - is to investigate how this technology with the proposed optical readout can be effective to reach this longer term goal. 
  
Moreover, the underground commissioning of such detector (in absence of passive neutron shielding) will be able to provide a precise seasonal, spectral and directional measurement of the LNGS fast and thermal neutron flux, of great interest for any current or future DM experiment in this location. After commissioning, with the proper water or polyethilene shielding, CYGNO will serve as demonstrator for the development of a large scale directional DM detector at the 30 m$^3$ scale. 

\section{CYGNO 1 m$^3$ goals}
The aim of the CYGNO project is to prove the capability of a high resolution gaseous TPC with sCMOS optical readout for near future direct Dark Matter searches at low 1-10 GeV WIMP masses down to and beyond the Neutrino Floor. For this reason, 1 m$^3$ demonstrator to be installed at LNGS (phase-1) aims to show the following performances and capabilities:
\begin{itemize}
\item O(keV) energy threshold on nuclear recoil tracks
\item 3D tracking reconstruction with head-tail determination
\item 3D detector fiducialization
\item Electron rejection power of 10$^5$
\item Complete background rejection thank to the very good particles detection and identification
\item Operate light target, such as He based gas mixture
\item High sensitivity in the 1-10 GeV WIMP mass region.
\end{itemize}

When achieved this results, the plan are to move to phase-2 for the construction of a 30-100 $m^3$  detector with a competitive sensitivity for the DM research and to reach the neutrino floor. Simultaneously, in other parts of the world (Australia, Japan, UK, China and US) there are similar proposals to build a distributed observatory of solar and galactic particles detectors, increasing total sensitivity and reducing systematic errors.

\begin{figure}[!ht]
\centering
\includegraphics[width=0.9\linewidth, angle=0]{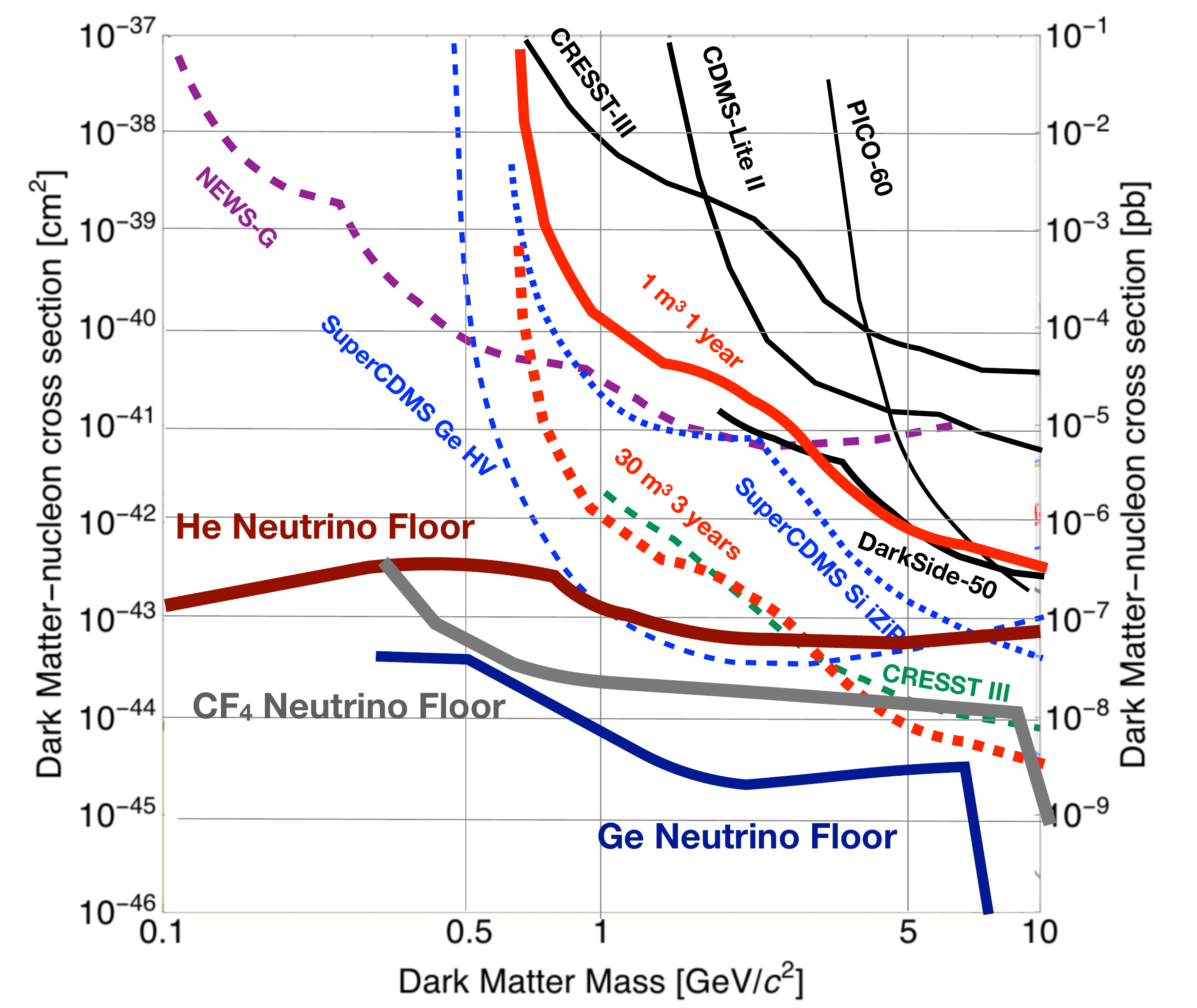}
\caption{Preliminary expected phase-1 and phase-2 Spin Independent 90\%C.L. exclusion limit  for a 1 m$^3$ (30 m$^3$) 80:20 He:CF$_4$  at atmospheric pressure with 1 keV threshold on He, 2 keV on C and 3 keV on F in 1 (3) year livetime and zero background, compared to current best limits (full lines) and expected sensitivities of future experiments (dashed lines), including CYGNUS-TPC. He Neutrino floor is extracted from \cite{Boehm:2018sux}, while CF$_4$, Ge and Xe floors from \cite{Ruppin:2014bra}.}
\label{fig:limits} 
\end{figure}

\begin{figure}[ht]
\centering
\includegraphics[width=0.9\linewidth, angle=0]{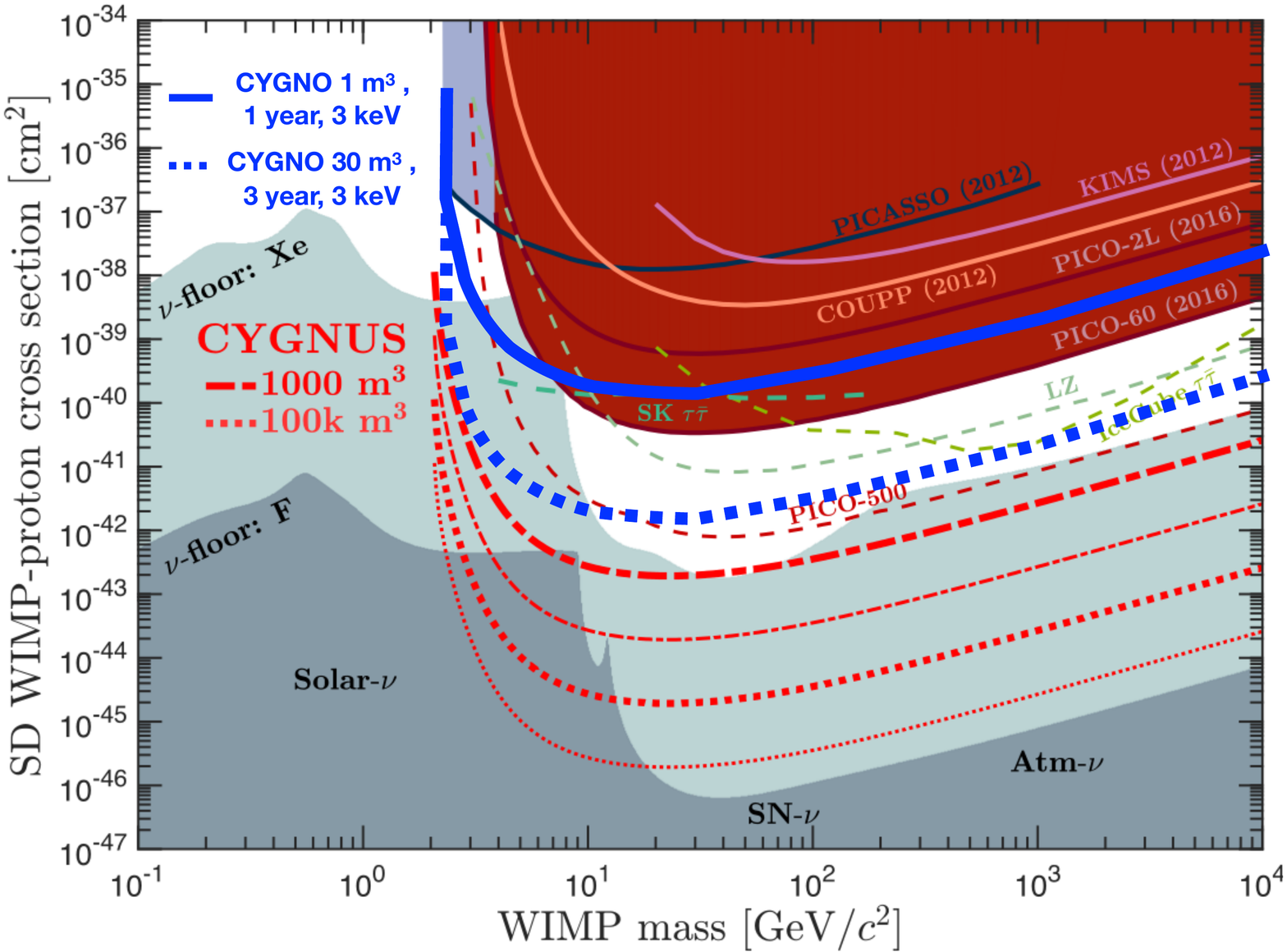}
\caption{Preliminary expected phase-1 and phase-2 Spin Dependent 90\%C.L. exclusion limit for a 1 m$^3$ (30 m$^3$) 80:20 He:CF$_4$  at atmospheric pressure with 1 keV threshold on He, 2 keV on C and 3 keV on F in 1 (3) year livetime and zero background, compared to current best limits (full lines) and expected sensitivities of future experiments (dashed lines), including CYGNUS-TPC.}
\label{fig:limits2} 
\end{figure}

The expected Spin Independent (Fig.\ref{fig:limits}) and Spin Dependent (Fig.\ref{fig:limits2}) 90\%C.L. exclusion limit for a 1 m$^3$ (30 m$^3$ dashed line) He CF$_4$ (80:20) are reported in . 

Up to now, the phase-0 of the CYGNO project has been funded by National Institute of Nuclear Physics, in order to produce a Technical Design Report by the end of 2019, and is supported by many international researchers and institution. In the mean time, the study of innovative and application of technique based on Negative Ion drift, and minority carrier are under study thanks to the support of the a recent grant from European Community.  

\section*{Conclusion}
CYGNO is a demonstrator of a promising readout technology applied to the detection of directional DM search. It aims to make a breakthrough in the usage of gaseous TPC-based detectors for directional DM exploiting from one side the advantage of low nose high granularity sCMOS sensors and from the other side the low mass target, such as He one, reaching an high sensitivity in the 1-10 GeV WIMP mass region. Moreover, CYGNO, as part of the CYGNUs international collaboration, will be able to increase the overall knowhow for the design and constriction of gas-only and low density large detector and able to approach the neutrino floor sensitivity.

\end{document}